\documentclass[10pt,nocopyrightspace,preprint,numbers]{sigplanconf}

\usepackage{amsmath}
\usepackage{url}
\usepackage[utf8]{inputenc}
\usepackage{graphicx}

\begin{document}

\setlength{\pdfpageheight}{\paperheight}
\setlength{\pdfpagewidth}{\paperwidth}

\conferenceinfo{PLDI '17}{Month d--d, 20yy, City, ST, Country}
\copyrightyear{2017}
\copyrightdata{978-1-nnnn-nnnn-n/yy/mm}
\copyrightdoi{nnnnnnn.nnnnnnn}

\title{A Synthesizing Superoptimizer}

\authorinfo{Raimondas Sasnauskas}
           {SES Engineering}
           {raimondas.sasnauskas@ses.com}
\authorinfo{Yang Chen}
           {Nvidia, Inc.}
           {yangchen@nvidia.com}
\authorinfo{Peter Collingbourne}
           {Google, Inc.}
           {pcc@google.com}
\authorinfo{Jeroen Ketema}
           {Embedded Systems Innovation by TNO}
           {jeroen.ketema@tno.nl}
\authorinfo{Gratian Lup}
           {Microsoft, Inc.}
           {gratilup@microsoft.com}
\authorinfo{Jubi Taneja}
           {University of Utah}
           {jubi@cs.utah.edu}
\authorinfo{John Regehr}
           {University of Utah}
           {regehr@cs.utah.edu}

\maketitle

\newcommand{\cL}{{\cal L}}
\newcommand*{\var}[1]{{\texttt{\%#1}}}
\newcommand{\tool}{Souper}

\begin{abstract}
If we can automatically derive compiler optimizations, we might be
able to sidestep some of the substantial engineering challenges
involved in creating and maintaining a high-quality compiler.
We developed \tool{}, a synthesizing superoptimizer, to see how far
these ideas might be pushed in the context of LLVM\@.
Along the way, we discovered that \tool's intermediate representation
was sufficiently similar to the one in Microsoft Visual C++ that
we applied \tool{} to that compiler as well.
Shipping, or about-to-ship, versions of both compilers contain
optimizations suggested by \tool{} but implemented by hand.
Alternately, when \tool{} is used as a fully automated optimization
pass it compiles a Clang compiler binary that is about 3\,MB (4.4\%)
smaller than the one compiled by LLVM\@.
\end{abstract}

\section{Introduction}

An ahead-of-time compiler is typically structured as a frontend, a
collection of optimizations, and a backend.
The optimizations in the ``middle-end'' of the compiler are numerous,
time-consuming to develop, hard to get right, and accrete assumptions
about costs that are difficult to excise as hardware platforms evolve.
An alternate strategy for implementing some parts of a middle-end is
to use a superoptimizer: a program that looks at the code being
compiled and uses a search procedure, a cost function, and an
equivalence verifier to automatically discover better (or even
optimal) code sequences.
The idea dates back at least 37~years~\cite{Fraser79}, and has been
the subject of dozens of papers since then.
We created \tool, a synthesizing superoptimizer that automatically
derives novel middle-end optimizations; it was originally designed for
LLVM~\cite{Lattner04} but we have also used it to find new
optimizations for the Microsoft Visual C++ compiler.

Several trends convinced us that it was time to write a new
superoptimizer.
There has been increased pressure on compiler
developers due to the adoption of higher-level programming languages and a
proliferation of interesting hardware platforms.
SAT and SMT solvers continue to improve; they are already more than
capable of discovering equivalence proofs necessary to verify compiler
optimizations involving tens to hundreds of instructions.
Solvers are also a key enabler for program synthesis, which supports
the discovery of new optimizations that are out of reach for naïve
search.
Finally, verified compilers appear to be much more difficult to extend
than are traditional compilers.
Though we have not yet done so, a natural extension of
superoptimization research would be to use a proof-producing solver to
greatly reduce the effort involved in incorporating new optimizations
into a proved-correct compiler~\cite{Leroy09, Mullen16}.
In summary, it appears to be a good time to re-evaluate some aspects
of compiler implementation, including how middle-end optimizers are
constructed.

Our contributions include the design and implementation of \tool{}, a
synthesis-based superoptimizer for a domain-specific intermediate
representation (IR) that resembles a purely functional,
control-flow-free subset of LLVM IR\@.
\tool{} implements an extension of Gulwani et al.'s synthesis
algorithm~\cite{Gulwani11}, allowing it to synthesize
LLVM's bitwidth-polymorphic instructions.

\tool{} has two intended use cases.
First, its results can be turned into actionable advice for compiler
developers; both LLVM and Microsoft compiler developers have
implemented optimizations suggested by \tool{}.
Second, \tool{} can run as an LLVM optimization pass, ensuring that
its code improvements can be exploited by other passes, such as
constant propagation and dead code elimination.
When building LLVM itself, \tool{} discovers about 7,900 distinct
optimizations, many of which cannot be performed by LLVM on its own,
and applies them a total of 85,000 times.
An \tool-optimized Clang-3.9 binary is almost 3\,MB (4.4\%) smaller
than one built without \tool, though it is also about 2\% slower.
(We do not yet know why; in this configuration, the only optimizations
performed by \tool{} were replacing variables by constants. This
should not hurt performance. In fact, versions of \tool{} based on
earlier versions of LLVM did get speedups in this case.)
Although the initial compilation of a program using \tool{} is often
$5\times$ to $25\times$ slower than optimized compilation with LLVM,
\tool{}'s discoveries are cached and subsequent compilations have much
lower overhead.
For example, the time for \tool{} with a warm cache to compile
LLVM on our test machine is about nine minutes, as opposed to
about eight minutes without \tool{}.

\section{\tool{} Design and Implementation}

The middle-end of a compiler is an exercise in compromises.
Much high-level language information, especially about types and about
structured control flow, has been thrown away.
At the same time, the target platform is frustratingly out of reach,
and it is difficult or impossible to take advantage of processor-level
tricks such as conditional execution and special-purpose instructions.

So, why have we developed a new middle-end superoptimizer?
First, LLVM IR is the narrow waist in a large and growing ecosystem of
frontends and backends; improvements made at this level can benefit
many projects and billions of end users (via, for example, Android).
Second, \tool{} excels at generating constants, particularly for
Boolean valued variables that are used to control branches.
Constants ripple through the rest of the middle-end and the full benefits are
not realized until constant propagation, dead code elimination, and other
optimization passes have exploited them.
Generating constants in the backend would leave these benefits on the
table.
Third, the SSA form that many modern compilers use in their middle
ends is effectively a functional programming language~\cite{Appel98}
that is highly amenable to automated reasoning techniques.

\subsection{An IR for Superoptimization}
\label{sec:ir}

\begin{figure}
\includegraphics[width=\columnwidth]{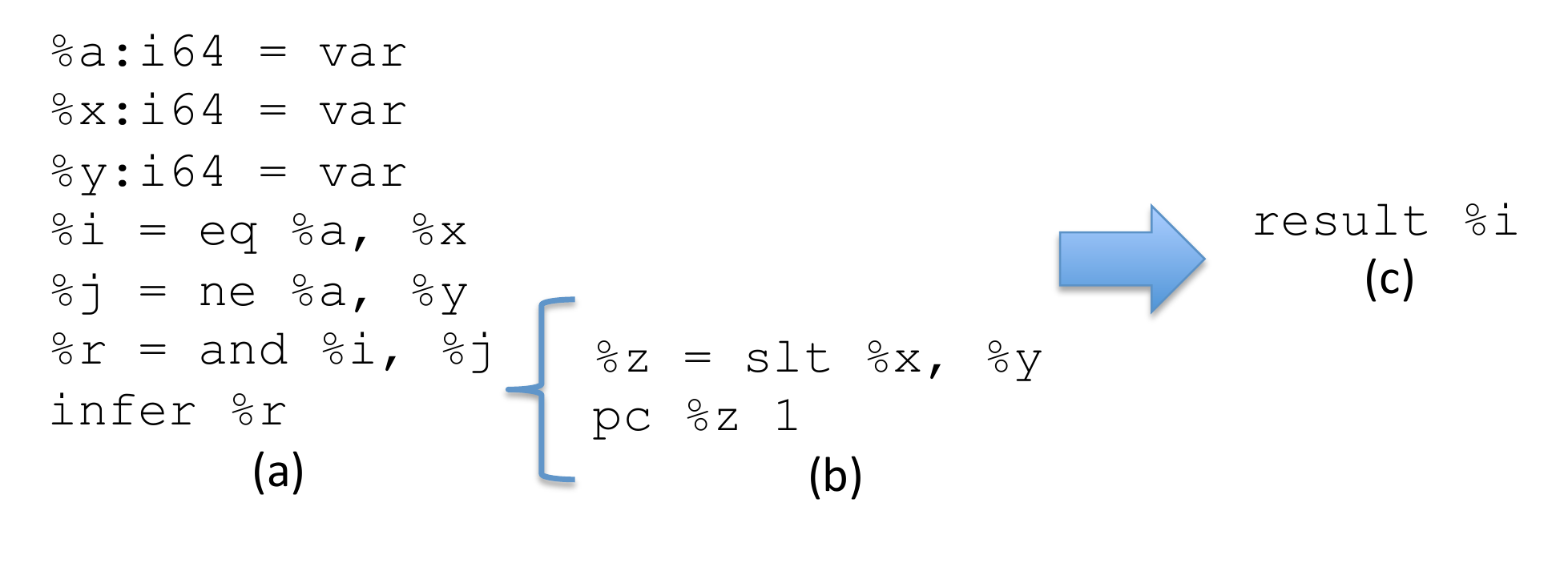}
  \caption{A simple \tool{} optimization. The left-hand side in (a)
    does not optimize.  Adding the path condition in (b) allows
    \tool{} to synthesize the right-hand side (c).}
  \label{fig:example1}
\end{figure}

\tool{}'s basic abstraction is a directed acyclic dataflow graph.
Operations closely follow those in the LLVM
IR,\footnote{\url{http://llvm.org/docs/LangRef.html}} though \tool{}'s
IR is purely functional.
\tool{} has 51 instructions, all derived from equivalents in the
integer, scalar subset of the LLVM instruction set.

The order of statements in \tool{} IR matters only in that a value
may not be referenced before it is defined.
An operand is either an integer constant or a value.
Integer constants have explicit widths; for example, the largest
signed 8-bit value is \texttt{127:i8}.
Constants may be either signed or unsigned, but this is only a
notational convenience: \tool{}, like LLVM and like processors,
associates signedness information with instructions rather than with
variables.
The only data types are bitvector, tuple of bitvector, and a special
block type.
Most operations are polymorphic with respect to bit width, and a few
type constraints are enforced:
\begin{itemize}
\item
  All bitvectors are at least one bit wide
\item
  Operand widths for instructions such as \texttt{add} must match
\item
  Comparison instructions return a single bit, and the first argument
  to a \texttt{select} instruction---LLVM's version of the ternary
  \texttt{?:} operator in C and C++---must be one bit wide
\item
  The \texttt{sext} (sign extend) and \texttt{zext} (zero extend)
  instructions must extend their argument by at least one bit
\item
  The \texttt{trunc} (truncate) instruction must reduce the bitwidth of its
  argument by at least one bit
\item
  Checked math operations such as \texttt{uadd.with.overflow} return
  a tuple consisting of the possibly-wrapped result and a bit
  indicating whether overflow occurred
\item
  The \texttt{extractvalue} instruction, like its LLVM counterpart,
  extracts a bitvector from a tuple
\end{itemize}
\tool{} does not perform type inference, but widths can be omitted in
most situations where they are obvious from context.

\subsection{Left-Hand Sides, Right-Hand Sides, and Path Conditions}

\tool{} does not have control flow, though it does provide two
constructs for representing dataflow facts learned from control flow
in the program being optimized.
%
%
For example, Figure~\ref{fig:example1}(a) shows the \emph{left-hand
  side} (LHS) of a potential optimization: the \texttt{infer} keyword
indicates the root node that \tool{} is being asked to optimize the
computation of.
Here \var{r} is a Boolean value that is true when the 64-bit input
\var{a} is equal to \var{x} and unequal to \var{y}.
In this case, \tool{} is unable to synthesize a better way to compute
\var{r}.

Now assume this left-hand side is augmented with the fact that \var{x}
is less than \var{y} using a signed comparison.
In \tool{}, such a fact is encoded as a path condition---an assertion
that a variable holds a particular value, that typically comes from
executing through a conditional branch in a program---that is shown in
Figure~\ref{fig:example1}(b).
Given this additional information, \tool{} can prove that \var{r} is
true if \var{a} equals \var{x}, and therefore the second comparison
and the \texttt{and}-instruction are unnecessary and can be dropped.
\tool{} represents a synthesis result as a \emph{right-hand side}
(RHS) containing the \texttt{result} keyword.
A right-hand side is a DAG that may refer to values on its
corresponding left-hand side.
A RHS is useful if it is cheaper to compute than its LHS\@.

A left-hand and right-hand side can be concatenated to create a
\emph{complete optimization}.
When presented with such an optimization, \tool's role is not to
synthesize, but rather to verify the correctness of the
optimization---a relatively straightforward job.

\subsection{Exploiting Correlated Phi Nodes}
\begin{figure*}
\begin{minipage}[b]{0.30\textwidth}
{\small\begin{verbatim}
int
f(bool cond, int z) {
  int x, y;
  if (cond) {
    x = 3 * z;
    y = z;
  } else {
    x = 2 * z;
    y = 2 * z;
  }
  return x + y;
}
\end{verbatim}}
\centering (a)
\end{minipage}
\begin{minipage}[b]{0.40\textwidth}
{\small\begin{verbatim}
define i32 @f(i1 %0, i32 %1) {
  br i1 %0, label %3, label %5

label %3:
  %4 = mul nsw i32 %1, 3
  br label %8

label %5:
  %6 = shl nsw i32 %1, 1
  %7 = shl nsw i32 %1, 1
  br label %8

label %8:
  %.07 = phi i32 [ %4, %3 ], [ %6, %5 ]
  %.0 = phi i32 [ %1, %3 ], [ %7, %5 ]
  %9 = add nsw i32 %.07, %.0
  ret i32 %9
}
\end{verbatim}}
\centering (b)
\end{minipage}
\begin{minipage}[b]{0.33\textwidth}
{\small\begin{verbatim}
%0 = block 2
%1:i32 = var
%2:i32 = shlnsw %1, 1:i32
%3:i32 = phi %0, %1, %2
%4:i32 = mulnsw 3:i32, %1
%5:i32 = phi %0, %4, %2
%6:i32 = addnsw %3, %5
infer %6
\end{verbatim}}
$\Rightarrow$
{\small\begin{verbatim}
%7:i32 = shl %1, 2:i32
result %7
\end{verbatim}}
\centering (c)
\end{minipage}
\caption{A C++ function that can be optimized to ``\texttt{return z <<
    2}'' (a), its representation in LLVM~IR (b), and a corresponding
  \tool{} LHS and RHS (c).  The first line of the LHS defines a block
  value with two predecessors; each phi node attached to this block
  will have two regular value arguments.  Then, the first argument to
  each phi node, \texttt{\%0}, establishes that they are correlated:
  since they come from the same basic block, they must make the same
  choice. Without information about the correlation between the phi
  nodes, this optimization cannot be synthesized.}
\label{fig:phi}
\end{figure*}

When control flow in LLVM code arrives at the start of a basic block,
each phi node~\cite{Cytron91} selects the value from its argument list
that corresponds to the block's immediate predecessor.
Lacking control flow, \tool{} preserves information about
correlated phi nodes using values of \emph{block type}.
Figure~\ref{fig:phi} shows an optimization that can only be performed
using the knowledge that two phi nodes make correlated choices.

\subsection{Reasoning About Incoming Control Flow}
\begin{figure*}
\begin{minipage}[b]{0.25\textwidth}
{\small\begin{verbatim}
unsigned
g(unsigned a) {
  switch (a % 4) {
  case 0:
    a += 3;
    break;
  case 1:
    a += 2;
    break;
  case 2:
    a += 1;
    break;
  }
  return a & 3;
}
\end{verbatim}}
\centering (a)
\end{minipage}
\begin{minipage}[b]{0.45\textwidth}
{\small\begin{verbatim}
define i32 @g(i32 %0) {
label %1:
  %2 = urem i32 %0, 4
  switch i32 %2, label %9 [
    i32 0, label %3
    i32 1, label %5
    i32 2, label %7
  ]

label %3:
  %4 = add i32 %0, 3
  br label %9

label %5:
  %6 = add i32 %0, 2
  br label %9

label %7:
  %8 = add i32 %0, 1
  br label %9

label %9:
  %.0 = phi i32 [ %0, %1 ], [ %8, %7 ],
                [ %6, %5 ], [ %4, %3 ]
  %10 = and i32 %.0, 3
  ret i32 %10
}
\end{verbatim}}
\centering (b)
\end{minipage}
\begin{minipage}[b]{0.33\textwidth}
{\small\begin{verbatim}
%0 = block 4
%1:i32 = var
%2:i32 = urem %1, 4:i32
%3:i1 = ne 0:i32, %2
%4:i1 = ne 1:i32, %2
%5:i1 = ne 2:i32, %2
blockpc %0 0 %3 1:i1
blockpc %0 0 %4 1:i1
blockpc %0 0 %5 1:i1
blockpc %0 1 %2 2:i32
blockpc %0 2 %2 1:i32
blockpc %0 3 %2 0:i32
%6:i32 = add 1:i32, %1
%7:i32 = add 2:i32, %1
%8:i32 = add 3:i32, %1
%9:i32 = phi %0, %1, %6, %7, %8
%10:i32 = and 3:i32, %9
infer %10
\end{verbatim}}
$\Rightarrow$
{\small\begin{verbatim}
result 3:i32
\end{verbatim}}
\centering (c)
\end{minipage}
\caption{A C or C++ function that can be optimized to ``\texttt{return 3}''
  (a), its representation in LLVM~IR (b), and the corresponding \tool{}
  LHS and RHS
  (c). The \tool{} IR has no control
  flow, but rather uses \texttt{blockpc} instructions to represent
  dataflow facts derived from converging control flow. These are
  crucial in giving \tool{} the information that it needs to
  synthesize the optimization.}
\label{fig:blockpc}
\end{figure*}

Consider the C function in Figure~\ref{fig:blockpc}(a).
In each case of the switch statement, \tool{} can use path condition
information extracted from control flow.
For example, the \tool{} LHS corresponding to the value of \texttt{a}
at the end of the first case (equivalently, corresponding to
\texttt{\%4} in Figure~\ref{fig:blockpc}(b)) is:

{\small\begin{verbatim}
%0:i32 = var
%1:i32 = urem %0, 4:i32
pc %1 0:i32
%2:i32 = add 3:i32, %0
infer %2
\end{verbatim}}

This asserts that at this program point, the remainder is zero.
The problem is that when we look at the program point where
information about the remainder values is actually useful---at the end
of \texttt{g}---we no longer have any path conditions, their
information has been lost as the control flow edges come together.
In other words, while path conditions capture useful information as
conditional branches are executed, they cannot represent facts about
convergent control flow.

To support reasoning about merged paths, \tool{} has a
\texttt{blockpc} construct: a reverse path condition tied to a value
of the block type.
The \tool~IR in Figure~\ref{fig:blockpc}(c) has six blockpcs, all tied
to block \texttt{\%0}.
The first, \texttt{blockpc \%0 0 \%3 1:i1}, asserts that when any phi
node tied to block \texttt{\%0} chooses its zeroth input, then
$\texttt{\%3} = 1$.
This fact, in turn, implies that \texttt{\%2}, the remainder, is not
equal to zero.
The next two blockpcs establish that the remainder is not one or two.
Together, these three conditions constrain the implicit default case
of the switch operator, which is the case where the remainder is three.
The next three blockpcs respectively constrain the three explicit
switch cases.
The overall effect is that \tool{} is able to reassemble information
learned in the switch cases in such a way that it is feasible to
derive the overall optimization.

\subsection{Extracting \tool{} IR from LLVM IR}

Since \tool{} has its own IR, it can run as a standalone tool.
However, we commonly want to convert LLVM IR into \tool{} IR on the fly, in
order to look for optimizations that can be applied to programs that we care
about.
For this purpose \tool{} uses LLVM's APIs to access the in-memory
representation of its IR\@.

\tool's \emph{extractor} scans each instruction in an LLVM module---a collection
of functions roughly equivalent to a C or C++ compilation unit---looking for
those that return integer-typed values.
Each such instruction leads to the construction of, and is the root
of, an \tool{} left-hand side.
The rest of the left-hand side is constructed by recursively following backwards
dataflow edges from this root instruction.
As the backwards traversal passes conditional branches and phi nodes
it adds path conditions and blockpcs.

To remain sound in the presence of loops, \tool{} must not extract any program
point more than one time in a single left-hand side.
In the vast majority of cases, LLVM's built-in loop detection suffices, but LLVM
does not detect irreducible loops, which occur rarely but which \tool{} must
detect on its own.
\tool{}'s backwards traversal is also stopped when it reaches a value
that comes from another function, a load from memory, or an
instruction that \tool{} lacks a model for, such as a floating point
or vector instruction.

Extraction is accompanied by canonicalization; arguments to
commutative operations are sorted; \tool{} canonicalizes away as many
comparison instructions as possible, for example turning greater-than
into less-than and swapping the operands; BitCast, PtrToInt, and
IntToPtr instructions simply pass on their bitvector values, throwing
away LLVM-level type information; GetElementPointer, LLVM's struct and
array element address generation instruction, is reduced to adds and
multiplies.

\subsection{Intrinsics}

We implemented 10 LLVM intrinsics as \tool{} instructions:
\begin{itemize}
\item
  Six operations that perform integer math while checking for
  overflow: there are signed and unsigned variants of add, subtract,
  and multiply
\item
  \texttt{ctpop}: Hamming weight
\item
  \texttt{bswap}: byte swap
\item
  \texttt{cttz} and \texttt{ctlz}: count trailing and leading zeroes
\end{itemize}

Combining several of these intrinsics, here \tool{} proves that the
Hamming weight of an arbitrary 64-bit value, multiplied by its number
of trailing zeroes, cannot overflow.

{\small\begin{verbatim}
%0:i64 = var
%1:i64 = ctpop %0
%2:i64 = cttz %0
%3 = smul.with.overflow %1, %2
%4 = extractvalue %3, 1
infer %4
\end{verbatim}}
$\Rightarrow$
{\small\begin{verbatim}
result 0:i1
\end{verbatim}}

The prevalence of these instructions varies across programs but, for
example,
UBSan\footnote{\url{http://clang.llvm.org/docs/UndefinedBehaviorSanitizer.html}}
inserts a large number of overflow checks into a program as it is
being compiled, and we would like to remove as many of these as
possible.

\subsection{Verifying Optimizations and Supporting Undefined Behavior}

\tool{} can be used to verify an optimization that was derived by hand
or synthesized previously---perhaps by an untrusted solver or
untrusted organization.
\tool{} is capable, for example, of proving equivalence of the first three
implementations of Hamming weight listed on the Wikipedia
page,\footnote{\url{https://en.wikipedia.org/wiki/Hamming_weight}} after they
are compiled to LLVM and then extracted into \tool{}.
To handle the fourth implementation, \tool{} would need to completely unroll the
loop, and to handle the fifth, it would need to model loads from a lookup table.

Verification follows the standard technique: \tool{} asks the solver
whether there exists any valuation of the inputs that causes
the left-hand and right-hand sides of the optimization to be unequal.
If this query is unsatisfiable, equivalence has been proved and
the optimization is sound.
If the query is satisfiable, a counterexample has been discovered and
it is presented to the user.
This is all fairly straightforward unless undefined behavior is
involved.

LLVM has three kinds of undefined behavior.
First, immediate undefined behavior, triggered by actions such as
dividing by zero or storing to an out-of-bounds memory location.
Second, an \emph{undef} value that stands for an indeterminate
register or memory location: it can return any value of its type.
Third, a \emph{poison} value that is more powerful than undef:
instructions other than \texttt{phi} and \texttt{select} return poison
if any input is poison.
Phi and select only return poison if the selected input is poison.
A poison value triggers true undefined behavior if it reaches a
side-effecting operation.

\tool{} does not model undef.
This is an acceptable approximation since undef usually occurs in the
context of uninitialized memory, and \tool{} has no model for memory.
When \tool{} encounters an explicit undef value while extracting
LLVM~IR into \tool{}~IR, it is conservatively modeled as zero.
Poison values are modeled by noting that any poison value in an \tool{}
expression will propagate to the root and trigger true undefined
behavior unless it is stopped by reaching the not-chosen input of a
phi or select instruction.
\tool{} models this behavior faithfully: each phi or select is
accompanied by an explicit path condition that only permits undefined
behavior to propagate via its selected branch.
The only immediate undefined behavior modeled by \tool{} is
divide-by-zero; \tool{} simply asserts to the solver that this does
not happen.

Some LLVM instructions have optional flags that add undefined behaviors.
For example, \texttt{add} is, by default, always defined, but
\texttt{add nsw} is undefined when signed overflow occurs, \texttt{add
  nuw} is undefined when unsigned overflow occurs, and \texttt{add nsw
  nuw} is undefined upon either kind of overflow.
\tool{} does not have flags, but rather provides separate instructions
corresponding to these different behaviors: \texttt{add},
\texttt{addnsw}, \texttt{addnuw}, and \texttt{addnswnuw}.
Similarly, the \texttt{exact} flag for division and right-shift, which
makes operations with remainders undefined, is modeled by supporting
\texttt{sdivexact} as well as \texttt{sdiv} and \texttt{ashrexact} as
well as \texttt{ashr}.

\subsection{Synthesizing Optimizations}

The essence of synthesis is finding a cost-minimizing solution to an
exists-forall formula.
In other words, we want to prove that there exists a way to connect up
a collection of instructions such that, for all inputs, the resulting
RHS behaves the same as its LHS\@.
Furthermore, the synthesized RHS should be the cheapest among all
that satisfy the behavioral requirement.

Given an equivalence checker, an algorithmically simple way to
implement synthesis is to enumerate, in order of increasing cost, all
RHSs, accepting the first one that passes the check.
In practice, this algorithm fails to produce results within an
acceptable amount of time when the cheapest RHS either contains
non-trivial constants or requires more than a handful of instructions.
To produce results in a more performant fashion, \tool{} uses
an improved version of the CEGIS (counterexample guided inductive
synthesis) algorithm developed by Gulwani et~al.~\cite{Gulwani11}.

CEGIS avoids exhaustive search and also avoids producing queries that
contain difficult quantifiers.
Rather, given a collection of instructions, it formulates a query
permitting all possible producer-consumer relations between
instructions, with the position of each instruction being represented
as a line number.
This query is satisfiable if there exists a way to connect the
instructions into a RHS that is equivalent to the LHS for just the
satisfying input.
If this is the case, the instructions and constants in the RHS can be
reconstructed from the model provided by the solver.
Then, in a second step, CEGIS asks the solver if the straight-line
program on the RHS is equivalent to the LHS for all possible inputs.
If so, synthesis has succeeded.
If not, constraints are added to prevent the solver from realizing
this particular circuit a second time, and the process is restarted.
In the worst case CEGIS is no better than naïve search, but in
practice it performs well, and has been shown to be capable of
generating RHSs of substantial complexity.
\tool{} wraps CEGIS in a second loop that constrains the size of the
synthesized RHS: it first attempts to synthesize a zero-cost RHS (no
new instructions are generated---the RHS may only use a constant or an
input), then a cost-one RHS (one instruction is generated), and so on.

Our improved CEGIS implementation can synthesize all \tool{} instructions
other than \texttt{phi}.
Using the \texttt{select} instruction, \tool{} can synthesize conditional
data paths.
Our algorithm is novel in that it can synthesize right-hand sides
composed of instructions that are polymorphic in the number of bits.
This requires \tool{} to augment the query with constraints that
enforce the bitwidth rules outlined in Section~\ref{sec:ir}.
A more difficult issue is that each instruction that is a candidate
for synthesis has a fixed bitwidth: there is no bitwidth-polymorphism
in the solver.
For each synthesis attempt, \tool{} starts with a default bitwidth:
the largest of any input and the target value.
Each instruction is instantiated at that width.
Next, for every input with a smaller bit width than the default bit
width, two extra components, a \texttt{zext} and a \texttt{sext}, are
instantiated that extend the input to the default width.
Finally, if the default width is larger than the output bit width of
the specification, \tool{} instantiates a \texttt{trunc} instruction
that truncates to the output width.
Similar transformations are applied to \texttt{select}'s conditional
input (\texttt{trunc}) and to the output of comparison instructions
 (\texttt{zext}).
These are heuristics and there is room for improvement.

\subsection{Validating \tool's Ability to Synthesize}

A synthesis implementation should be evaluated both in terms of
soundness and ability to synthesize a RHS when it exists.
We discuss soundness in more detail in Sections~\ref{sec:threats}
and~\ref{sec:validate}.
Regarding synthesis power, we ensured that \tool{} can synthesize a
subset of the ``Hacker's Delight'' optimizations from Gulwani et
al.~\cite{Gulwani11} where the instructions involved can be mapped to
\tool{}~IR in a one-to-one fashion and where the RHS is not too large
(P1--P17 and P19 out of P1--P25).
Our CEGIS implementation does not scale as well as Gulwani et al.'s,
we believe this is due to the extra synthesis components and constraints
relating to bitwidth rules.

We ensured that \tool{} can synthesize specific instances of every
optimization in Table~2 of Buchwald~\cite{Buchwald15}.
By ``specific instances,'' we mean that while Optgen can create a
general rule such as \texttt{\textasciitilde x + c} $\Rightarrow$ \texttt{(c - 1) - x}
that works for an arbitrary constant \texttt{c}, \tool{} must
rediscover the optimization for every bitwidth and every value of $c$
that appears in a program.

Finally, we validated \tool{}'s synthesis by ensuring that it can
solve discrete math problems.
Consider Mordell's equation, $y^2 = x^3 + k$, in the domain of natural numbers,
where $k$ is a constant.
We can encode a bounded version of this equation for $k = 7$, a case known
to not have any solutions, like this:
{\small\begin{verbatim}
%x:i32       = var
%y:i32       = var
%xsqr        = mulnuw %x, %x
%xcubed      = mulnuw %x, %xsqr
%ysqr        = mulnuw %y, %y
%xcubedplus7 = addnuw %xcubed, 7
%cmp         = eq %xcubedplus7, %ysqr
pc %cmp 1
infer %y
\end{verbatim}}

The ``nuw'' variants of multiplication and addition assert that unsigned integer
overflow is undefined, protecting us from undesirable wraparound effects.
The path condition asserts that the equation holds and the \texttt{infer} line
asks \tool{} for the value of $y$.
It fails to synthesize a RHS\@.

Similarly, \tool{} fails to synthesize a result when $k = 1$, because that case
has multiple solutions: $x=0$, $y=1$ and $x=2$, $y=3$.
On the other hand, for $k = 785$ there is a unique solution within the
bounds and \tool{} finds it:

{\small\begin{verbatim}
result 32146:i32
\end{verbatim}}
\noindent

\noindent
when asked to infer \texttt{\%x}:

{\small\begin{verbatim}
result 1011:i32
\end{verbatim}}

While we do not know of any use cases for solving Diophantine
equations inside an optimizing compiler, we are gratified to know that
the power is there should it be needed.

\subsection{Three Real Threats to Soundness}
\label{sec:threats}

We have observed miscompilations due to three causes other than the
obvious one (defects in \tool's implementation).
First, an incorrect LLVM optimization can turn a defined program into
an undefined one, but in a subtle way that is not exploited by an LLVM
backend.
There are known, long-standing bugs of this type in LLVM, and it is
difficult to get them fixed because their effects are difficult to
observe via end-to-end testing of the LLVM toolchain, and also because
the LLVM community has not reached consensus on exactly what the fix
should look like.
\tool{}, on the other hand, readily notices and exploits undefined
behaviors in order to perform computations more cheaply.
This is a highly unsatisfactory state of affairs, but fixing them
requires community-wide effort.
In the meantime, the problem can be mitigated by turning off some
problematic LLVM passes (SimplifyCFG, GVN, and InstCombine) or by
adjusting \tool{}'s LLVM semantics (using a command line option) to
prevent it from exploiting undefined behavior.

A second, closely related problem is that some applications
execute undefined behavior that happens to be benignly compiled by
LLVM\@.
In this case, the application, not LLVM, is wrong, but the end result
can be the same: \tool{} notices and exploits the undefined behavior
to break the application.
From this perspective, \tool{} might be viewed as a hostile
re-implementation of STACK~\cite{Wang13} that breaks programs instead
of providing helpful diagnostics.
This risk can be mitigated in two ways.
First, as above, UB exploitation in \tool{} can be disabled.
This is not a complete fix, but it does prevent some
particularly egregious problems.
Second, every UB exploited by \tool{} can be detected by LLVM's
undefined behavior
sanitizer.\footnote{\url{http://clang.llvm.org/docs/UndefinedBehaviorSanitizer.html}}
Developers should be using UBSan anyhow, since \tool{} is hardly the
worst problem faced by an undefined C or C++ program.

Third, when a solver is wrong, \tool{} will also be wrong.
One time we saw a program that had been optimized by \tool{}
misbehave, and the root cause was an incorrect result returned by the
Z3 solver.
We reported the bug and the Z3 developers rapidly fixed it.
%
Although we do not routinely check solvers against each other, we can
do so on demand since \tool's queries are in the SMT-LIB format
supported by most solvers.

\subsection{Validating Soundness}
\label{sec:validate}

Validating any particular optimization produced by \tool's synthesizer
is not difficult: each result that it creates can be verified by
\tool{}'s equivalence checker.
The equivalence checker is much simpler and we have more confidence in
it.

We validated \tool{}'s soundness in several ways.
First, we have stress-tested it using Csmith~\cite{Yang11} and also by
compiling significant programs, such as LLVM and SPEC CINT 2006,
that have good test suites.
Several SPEC benchmarks misbehave after being optimized by \tool{}
unless undefined behavior exploitation is disabled.
This is hardly surprising since some of these benchmarks are known to
execute undefined behaviors~\cite{Dietz15} and can even be broken by
GCC if the wrong optimization options are used.\footnote{The release
  notes for GCC~4.8 include this note: ``GCC now uses a more
  aggressive analysis to derive an upper bound for the number of
  iterations of loops using constraints imposed by language
  standards. This may cause non-conforming programs to no longer work
  as expected, such as SPEC CPU 2006 464.h264ref and 416.gamess. A new
  option, -fno-aggressive-loop-optimizations, was added to disable
  this aggressive analysis.''}
Some LLVM~3.9 test cases give the wrong answers when LLVM has been
optimized by \tool{}, even when undefined behavior exploitation is
turned off.
We believe these are due to undefined behaviors in LLVM, such as uses
of uninitialized storage, that cannot be easily mitigated just by
disabling undefined behavior exploitation in \tool{}.
We discuss this issue further in Section~\ref{sec:online}.
Second, we have validated \tool{} by looking at hundreds of
optimizations by hand, by posting optimizations on the web where LLVM
developers could see them, and by cross-checking them using
Alive~\cite{Lopes15}, which has an independent formalization of the
semantics of the LLVM instruction set.
In summary, we have made a good-faith attempt to validate \tool, but
it remains a research-quality optimizer that may well harbor exciting
bugs.

\subsection{Caching}
\label{sec:cache}

Since \tool{} must invoke a solver multiple times to synthesize an
optimization such as the ones shown in Figures~1--3, and since such
invocations often result in the solver being killed when a timeout
expires, we would like to amortize the costs of optimization
synthesis.
A reasonable solution is to cache the mapping of left-hand sides to
right-hand sides, including the null right-hand side indicating
failure to optimize.

\tool's first-level cache is a hash table in RAM, allowing very fast
lookups for commonly-occurring left-hand sides that are likely to be
encountered multiple times during a single compiler invocation.
\tool's second-level cache is Redis,\footnote{\url{http://redis.io/}} a fast,
networked key-value store.
%

\subsection{Implementation}
\label{sec:impl}

\tool{} is open source software and is implemented in about 9,500
lines of C++.
\tool{} currently links against LLVM~3.9.
\tool{}'s functionality can be invoked in a number of ways; there are
command-line tools for processing \tool{} IR and LLVM IR, and \tool{}
can be linked into a shared library that is dynamically loaded as an
LLVM optimization pass.
\tool{}'s LLVM pass registers itself using \texttt{EP\_Peephole}, an
extension point designed for peephole-like passes.
At the \texttt{-O2} or \texttt{-O3} levels, LLVM runs \tool{} five
times during compilation, giving it multiple opportunities to interact
with other optimization passes.
Finally, we have implemented a compiler driver \texttt{sclang} that is
a drop-in replacement for \texttt{clang} except that it loads the
\tool{} pass.
This makes it easy to build arbitrary software packages using \tool.

\tool{} uses Klee~\cite{Cadar08} as a library for emitting a query in SMT-LIB
format~\cite{Barrett10}.
By default, in order to model memory, Klee uses the theory of
bitvectors and arrays.
However, we observed that solvers such as Z3 can perform poorly in
this theory and also that \tool{} does not require a model for memory,
so we patched Klee to emit queries in the theory of quantified
bitvectors.

During the course of our work we ran into several degeneracies in Klee
triggered by, for example, large \tool{} LHSs with many phi nodes.
We fixed (and upstreamed) several issues, but in the end huge \tool{}
queries still performed poorly, triggering apparently exponential
behaviors.
We currently work around these issues by enforcing a limit on LHS
size: \tool{} simply drops any LHS that is over the specified limit,
which is configurable but defaults to 1\,KB of serialized IR\@.
This mechanism saves a lot of execution time while dropping a small
minority of queries.

\section{Experience with \tool{} as an Offline Optimization Generator}
\label{sec:offline}

The goal is to give compiler developers actionable advice about
missing optimizations.
To do this, someone uses \texttt{sclang} to extract \tool{} left-hand
sides from programs of interest.
Since \tool{} finds many more optimizations than developers could
reasonably implement, the crux is ranking them in such a way that
desirable optimizations are found early in the list.
We are not aware of any single best ranking function, but rather we
have created several such functions that can be used together or
separately.

\paragraph{Static profile count}
The first time a given LHS is encountered, it is stored in Redis.
Each subsequent time it is encountered, \tool{} simply increments a
static profile count, also stored in Redis, that is associated with
the optimization.
Thus, counts are automatically aggregated across multiple compiler
invocations.
Implementing optimizations with high static profile counts will lead
to optimizations that fire many times, presumably leading to benefits
in terms of code size.

\paragraph{Dynamic profile count}
To make programs faster instead of smaller, it is desirable to focus
on optimizations that execute many times dynamically, as opposed to
statically.
We support dynamic profiling in \tool{} by optionally instrumenting
each compiled code module to associate a 64-bit counter with each
(potentially) optimized code site.
These counters are atomically incremented each time the associated
site is reached, and then when the program is shutting down the total
values are added to dynamic profile counts in the Redis database.
Counts can be lost if a program crashes or
otherwise fails to execute its atexit
handlers.\footnote{\url{http://man7.org/linux/man-pages/man3/atexit.3.html}}

\paragraph{LHS complexity}
Since \tool{} does not have any default limit on the number of
instructions it extracts, LHSs may be large.
Developers are unlikely to want to implement recognizers for large
instruction patterns, and furthermore such patterns are unlikely to be
broadly applicable.
Therefore, it makes sense to suppress large LHSs when ranking
optimizations.
Alternatively, we have experimented with depth-limited extraction of
LHSs: this reduces the number of optimizations that can be
synthesized, but the optimizations that remain are much more likely to
be of interest to developers.

\paragraph{Benefit}
The benefit due to an optimization is the difference in cost between
the LHS and RHS\@.
So far we have employed only simple cost functions such as those that
count instructions (perhaps weighting expensive instructions such as
divides higher).
Doing better than this has proved difficult.
First, LLVM has many canonicalization rules which dictate that certain
IR forms are preferable over others; these rules are, unfortunately,
implicit and informal.
Second, despite the ``low level'' in LLVM, it is fairly high level,
delegating a lot of translation work to the backends, making it hard
to determine a cost model at the LLVM level.

\subsection{Improving LLVM}

There is one additional piece of the puzzle to solve before
synthesized superoptimizer results can be presented to developers in a
useful fashion: derived optimizations should be suppressed when LLVM
already knows how to perform them.
\tool{} gets LHSs that LLVM can optimize for two reasons.
First, \tool{} gets run five different times by the LLVM optimizer,
and during early phases many optimizations remain undone.
Also, even when LLVM is finished optimizing, there remain opportunities
that LLVM could optimize because it simply runs a fixed ordering of
passes, it does not run to fixpoint.
We could run LLVM's optimizer to fixpoint, but this would be pointless
because \tool{} would still extract many LHSs that LLVM can optimize,
because many performable optimizations are rejected by profitability
heuristics.
Typically, rewriting a code sequence into a better code sequence is
deemed unprofitable if values in the original sequence have external uses.
To solve this problem we translate each \tool{} LHS back into LLVM IR
and see if it optimizes.
Because no value in the translated LLVM code has external uses, this
is a fairly reliable way to avoid showing compiler developers
optimizations that have already been implemented.

In November 2014 and in July 2015 we presented ranked lists of
optimizations derived by \tool{} while compiling LLVM to the LLVM
community.
The 2014 results contained only synthesized integer values and the
2015 results had synthesized instructions as well.
This was a learning experience for us, particularly with respect to
the importance of having good LLVM-specific profitability estimation
methods.
For example, this optimization eliminates an LLVM instruction and
seems intuitively appealing:

{\small\begin{verbatim}
%0:i64 = var
%1:i64 = and 1:i64, %0
%2:i1 = ne 0:i64, %1
infer %2
\end{verbatim}}
$\Rightarrow$
{\small\begin{verbatim}
%3:i1 = trunc %0
result %3
\end{verbatim}}

However, it turns out that this optimization is undesirable by
convention, and in fact the LLVM optimizer will canonicalize the
\texttt{trunc} back to the two-instruction version.

We do not have a good way to quantify any improvements in LLVM's
optimizer that might have resulted from our work, but we do know that
some of \tool{}'s suggestions were implemented.
Here are some things that LLVM developers said:
\begin{itemize}
\item
  ``Cool! Looks like we do lots of provably unnecessary alignment checks. :)''
\item
  ``That’s a great post and really interesting data, thank you!''
\item
  ``I’m pretty sure I’ve fixed the most egregious cases of this going wrong with r222928.''
\item
  ``IIRC, the following commits are a direct/indirect result of using
  [deanonymized name of SSO]'' (followed by a list of seven commits)
\end{itemize}
Additionally, we seen some other commits that mention \tool{} as a
source, and we have watched a lot of optimizations with high profile
counts disappearing from \tool{}'s output as LLVM's optimizers have
become stronger over time.

We plan to continue periodically posting \tool-derived optimizations.
In particular, it will be exciting to extract LHSs from novel sources
of LLVM IR: Haskell, Swift, Rust, and UBSan output.
We expect that each of these will contain idioms that are frequent and
that are not well-optimized at present.
In particular, LLVM is known to be very weak at eliminating the
integer overflow checks inserted by UBSan.
Removing these is becoming more desirable as production code is deployed
with integer overflow checking turned on.%
\footnote{\url{http://android-developers.blogspot.com/2016/05/hardening-media-stack.html}}

\subsection{Improving Microsoft Visual C++}

The Visual C++ compiler IR is similar enough to LLVM that we were able
to extract a subset of it directly to \tool{} IR (for the workflow
described in this section, LLVM is not involved at all).
In this extractor we limited each LHS to five instructions and we did
not extract path conditions or blockpcs; this would be useful
future work.
We extracted LHSs for several major components of Windows, along with
their static profile counts.
In this section, we focus on LHSs extracted from the Windows kernel in
its x86-64 configuration.

\newcommand{\myrule}{\vskip0.15in\hrule\vskip0.10in}

\begin{figure*}

{\small

\begin{minipage}[t]{0.31\textwidth}

\myrule

{\begin{verbatim}
%0:i16 = var
%1:i16 = or %0, 5:i16
%2:i16 = and %1, 4:i16
infer %2
\end{verbatim}}
$\Rightarrow$
{\begin{verbatim}
result 4:i16
\end{verbatim}}

This example, and others like it, motivated us to implement a new
\emph{bit estimator} analysis in the Visual C++ compiler that identifies bits that are
provably zero or one, making this optimization and many related ones
easy to implement.

\myrule

{\begin{verbatim}
%0:i32 = var
%1:i32 = xor %0, 4294967295:i32
%2:i32 = and %1, 8:i32
infer %2
\end{verbatim}}
$\Rightarrow$
{\begin{verbatim}
%3:i32 = and 8:i32, %0
%4:i32 = xor 8:i32, %3
result %4
\end{verbatim}}

Instead of inverting the entire word and then isolating bit~3, we can
first isolate the bit and then invert it.
The resulting code contains a 1-byte immediate value instead of a
4-byte one.
(\tool{}'s cost function is not currently clever enough to capture
this fact---it was a matter of luck that this optimization was
synthesized.)

\myrule{
\begin{verbatim}
%0:i8 = var
%1:i8 = lshr %0, 3:i8
%2:i1 = eq %1, 0:i8
infer %2
\end{verbatim}}
$\Rightarrow$
{\begin{verbatim}
%3:i1 = ult %0, 8:i8
result %3
\end{verbatim}}

A value shifted right 3 times can be zero only if it is smaller than $2^3 = 8$.

\end{minipage}
\hspace{0.035\textwidth}
\begin{minipage}[t]{0.31\textwidth}
\myrule
{\begin{verbatim}
%0:i64 = var
%1:i64 = and %0, 31:i64
%2:i64 = add %1, 1:i64
%3:i1 = ult 32:i64, %2
infer %3
\end{verbatim}}
$\Rightarrow$
{\begin{verbatim}
result 0:i1
\end{verbatim}}

Adding 1 to a value in range $0..31$ cannot produce a result larger
than 32. The bit estimator can be used to implement the family of
optimizations suggested by this example.

\myrule
{\begin{verbatim}
%0:i64 = var
%1:i64 = mul %0, 2654435761:i64
%2:i64 = and %1, 1:i64
%3:i1 = ne %2, 0:i64
infer %3
\end{verbatim}}
$\Rightarrow$
{\begin{verbatim}
%4:i1 = trunc %0
result %4
\end{verbatim}}

Since multiplying two odd numbers always yields an odd number, the
result can be odd only if \%0 is odd to begin with.

\myrule
{\begin{verbatim}
%0:i32 = var
%1:i64 = zext %0
%2:i64 = udiv -1:i64, %1
%3:i1 = ule 4:i64, %2
infer %3
\end{verbatim}}
$\Rightarrow$
{\begin{verbatim}
result 1:i1
\end{verbatim}}

Dividing the largest unsigned 64-bit number by an unsigned 32-bit
number cannot produce a value that is smaller than $2^{32} - 1$.
We used the bit estimator to prove a lower bound on the value of \%2,
making is possible to fold the comparison to true.

\end{minipage}
\hspace{0.035\textwidth}
\begin{minipage}[t]{0.31\textwidth}

\myrule
{\begin{verbatim}
%0:i32 = var
%1:i1 = eq %0, 0:i32
%2:i32 = select %1, 2:i32, 1:i32
infer %2
\end{verbatim}}
$\Rightarrow$
{\begin{verbatim}
%3:i1 = ult %0, 1:i32
%4:i32 = zext %3
%5:i32 = shl 1:i32, %4
result %5
\end{verbatim}}

\tool{} excels at replacing \texttt{select} instructions, which
usually become CMOV instructions on x86/x86-64, with logical
operations.

\myrule
{\begin{verbatim}
%0:i1 = var
%1:i8 = select %0, 1:i8, 0:i8
%2:i1 = ne %1, 1:i8
%3:i32 = select %2, 40:i32, 20:i32
infer %3
\end{verbatim}}
$\Rightarrow$
{\begin{verbatim}
%4:i32 = zext %0
%5:i32 = ashr 40:i32, %4
result %5
\end{verbatim}}

Another \texttt{select} removal.

\myrule
{\begin{verbatim}
%0:i32 = var
%1:i32 = and %0, 131071:i32
%2:i32 = mul %1, 65536:i32
infer %2
\end{verbatim}}
$\Rightarrow$
{\begin{verbatim}
%3:i32 = shl %0, 16:i32
result %3
\end{verbatim}}

\tool{} is good at removing useless instructions.

\myrule
{\begin{verbatim}
%0:i64 = var
%1:i64 = add %0, -10445360463872:i64
%2:i64 = and %1, 4095:i64
infer %2
\end{verbatim}}
$\Rightarrow$
{\begin{verbatim}
%3:i64 = and 4095:i64, %0
result %3
\end{verbatim}}

Another useless instruction removed.

\end{minipage}

}

\vskip 0.3in

\caption{Representative optimizations suggested by \tool{} for LHSs
  extracted from the Windows kernel. We implemented generalized
  versions of these, and others, in the Visual C++ compiler, adding a
  total of 40 new optimizations.}
\label{fig:windows}
\end{figure*}

Out of 15,846 unique LHSs that were extracted, \tool's synthesizer
discovered a cheaper RHS for 935 of them within a one-minute timeout.
However, many of these optimizations were already supported---Visual
C++ has a large collection of SSA-level peephole optimizations---but
had not been performed since one or more of the values on the LHS had
uses not visible in the \tool{} IR\@.
We plan to implement a compiler-specific filter to automatically weed
out these undesired optimizations, like the one that we implemented
for LLVM, but we have not yet done so.
Out of the remaining optimizations, we implemented 40 that had
high static profile counts.
We manually rewrote each of these in a generic form, verified the
correctness of that form using Alive~\cite{Lopes15}, and then
implemented it in the Visual C++ compiler.
Figure~\ref{fig:windows} shows some representative examples.
Implementing the new optimizations required two new dataflow analyses
to be added to the Visual C++ compiler.
First, a \emph{bit estimator} that attempts to prove that individual
bits of values are zero or one.
Second, a \emph{demanded bits} analysis that attempts to show that
some bits of a value have no influence on the computation.
For example, if the only use of \texttt{x} is in computing \texttt{x |
  0xff}, the low eight bits of \texttt{x} are not demanded: their
value is irrelevant to the program.
Together, these analyses drive optimizations that remove useless
instructions.

The new optimizations that we implemented reduce the code size of the
Windows kernel by a few KB (about a 0.1\% savings) and the code size
of a Windows 10 build by 296\,KB for x86 and 316\,KB for x86-64 (about
a 0.02\% savings).
However, the effect of optimizations on code size is complicated:
there were some binaries in Windows that got larger; this was mostly
caused by more functions being inlined.

\section{Using \tool{} as an Online Optimizer}
\label{sec:online}

Every time an optimization suggested by \tool{} in its offline
capacity is implemented, \tool{} loses some of its power to optimize
online.
Nevertheless, we feel it is worthwhile to evaluate \tool{} online,
though its power is less than it would have been a few years ago.

When using \tool{} as an online compiler, we restrict it to
synthesizing constants, since it should nearly always be a win to
replace a value at the LLVM level with a constant.
Replacing an LHS in LLVM with instructions from an \tool{} RHS
introduces several performance-related difficulties that we have not
yet solved.
First, as discussed in Section~\ref{sec:offline}, estimating
profitability is difficult, especially in the presence of LLVM's
arcane canonicalization rules.
Second, \tool{} abstracts away the fact that some of the values in a
LHS are likely to have uses not visible in the \tool{} IR\@.
These uses will prevent instructions on the LHS from being eliminated
and may therefore dramatically reduce the benefit that is realized by
applying a particular optimization.
Third, naïvely replacing an LHS with a RHS can break SSA, since it is
not necessarily the case that the inputs to an LHS dominate the root of
a DAG of \tool{} IR\@.
Of course, this can be fixed up, but the fix is not free since it
forces values to be computed on code paths where they previously were
not computed.
In summary, there are significant challenges in automatically and
profitably implementing arbitrary \tool-derived optimizations that we
leave for future work.

\paragraph{Experimental setup}

For the experiments reported in this paper, we used an Intel i7-5820K
(3.3\,GHz, six-core Haswell-E) with 16\,GB of RAM running Ubuntu
14.04.
We configured the processor to use performance mode and disabled turbo
mode to minimize dynamic frequency scaling effects.
For the solver we used a current snapshot of Z3 version 4.5.1.
We used our patched Klee that emits queries in the theory of
quantified bitvectors rather than the theory of arrays.
Whenever possible, we ran compilations using all cores.
However, actual benchmarks were run on a single core of an otherwise
quiescent machine.

\paragraph{Optimizing LLVM}

The LLVM-3.9 compiler and its Clang frontend are, together, nearly
three million lines of C++.
We compiled them using \tool{}; this took about 88 minutes with a cold
cache and 14 minutes with a warm cache.
At the end of compilation, Redis was using 362\,MB of RAM and its
database dump file was 149\,MB.
In contrast, compilation without \tool{} took 13 minutes.
In both cases, the compilation was in release mode (full optimization,
no debugging symbols) with assertions enabled.

The \tool{}-optimized \texttt{clang} binary, which contains all of the
LLVM internals, is 64.3\,MB, in contrast with the non-\tool{}
binary which is 67.2\,MB: a savings of 2.9\,MB\@.
A lot of the code size savings appears to come from proving that
assertions cannot fire; when assertions are disabled (turned into nops
using the preprocessor), the \tool-optimized Clang binary is only
about 600\,KB smaller than the non-\tool{} version.
When used to perform optimized compiles of C++ code, the
\tool-optimized Clang is 2\% slower than the default release build of
Clang.
We suspect, but cannot (yet) prove that we are running afoul of an
unlucky combination of optimization interactions, since it should not
be the case that replacing values with constants (and often small
constants like 0 and 1 that are easy to materialize) results in worse
code generation.

When the test suite for the \tool{}-optimized LLVM/Clang is run, 92
out of about 17,000 test cases fail.
For every one of these that we inspected, LLVM/Clang executed
undefined behavior, which we believe to be the root cause of the
failing test cases, with \tool{} simply exposing the problem.
Our belief is not yet backed up by strong evidence, but we plan to work
with the LLVM developers to eliminate the undefined behaviors, and see
if the failures then go away.
If they do not, the evidence points to a defect in \tool{} that
we have been otherwise unable to locate.
We noticed that all of the \tool{} optimizations that triggered test
case failures involved blockpcs.
The 2\%-slower Clang executable described in this section was compiled
by \tool{} in a mode that did not extract blockpcs.

\paragraph{Is \tool{} acceptably fast?}

\begin{figure}
  \includegraphics[width=\columnwidth]{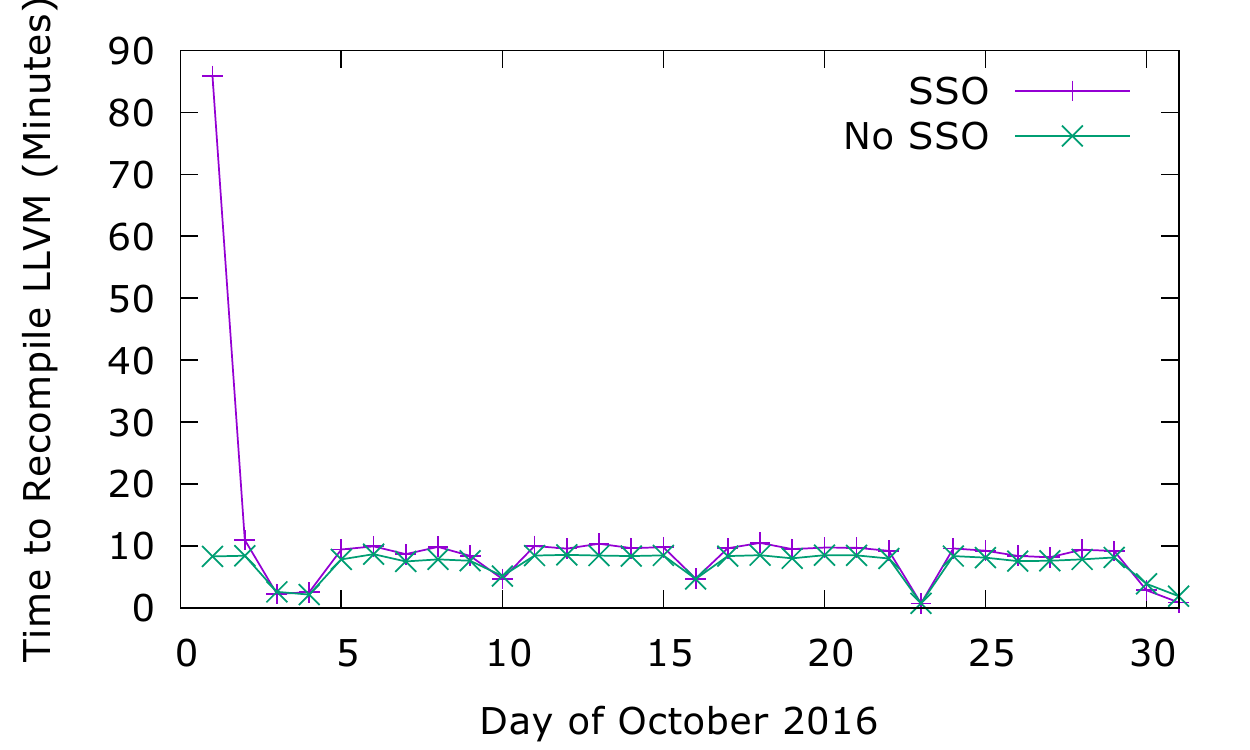}
  \caption{Time taken to incrementally compile the LLVM development
    trunk at the start of each day of October 2016, with and without
  \tool{}.}
  \label{fig:daily}
\end{figure}

We simulated the experience a developer using \tool{} might have by
incrementally compiling LLVM (without Clang) from its development
trunk at the start of each day in October 2016.
Build times are shown in Figure~\ref{fig:daily}.
For a developer not using \tool{}, the first build took a little over
eight minutes, and so did most subsequent builds, because on most days
the LLVM developers checked in patches to one or more widely-included
header files, forcing a near-total recompilation.
On a few days, no such changes were committed and incremental
compilation was fast; it required less than one minute on October~23
and~31.

For the developer using \tool{}, compilation on October~1 took about
86 minutes because the cache was cold and the solver had to be called
many times.
However, for the rest of the month, this developer could take
advantage of the fact that most of a large code base does not change
frequently, meaning that most \tool{} queries were satisfied by the
cache.
The \tool{} user's LLVM build was a little over a minute slower, on
average, than the non-\tool{} build, during October 2--31.

\tool{} has a mode where, instead of attempting to optimize every
integer-typed LLVM value, it only attempts to optimize 1-bit values.
In this mode, warm-cache compiles using \tool{} can be faster than
LLVM alone: enough code gets eliminated early in compilation to more
than make up for \tool{}'s execution costs.

\paragraph{Optimizing SPEC CINT 2006}

We compiled the C and C++ integer benchmarks from SPEC CPU 2006 using
\tool{}, with undefined behavior exploitation turned off.
A parallel build of the benchmarks using \tool{} took 26 minutes with
a cold cache, and this improved to 2~minutes 15~seconds when the cache
was warm.
In contrast, building the benchmarks without \tool{} took 1~minute
5~seconds.
At the end of the build, Redis was using 104\,MB of RAM and its
database dump file was 42\,MB\@.
Across all benchmarks, \tool{} looked at about 126,000 distinct LHSs
and had a total of about three million opportunities to optimize a
LHS\@.
However, it discovered only 898 distinct optimizations, and applied
them 2212 times.

The effect of \tool{} on code size was uneven: seven benchmarks (bzip,
perlbench, h264ref, gobmk, astar, hmmer, xalancbmk) became larger
while five (omnetpp, libquantum, gcc, sjeng, mcf) became smaller.
Total code size across all benchmarks was increased by about 2\,KB, out
of a total of about 12\,MB\@.
An increase in code size is anomalous since we are running \tool{} in
a mode where it only replaces LLVM values with constants.
The explanation is that LLVM's inlining heuristics, which can be
somewhat sensitive, are responding to \tool's improvements and making
different inlining decisions.
We verified this by disabling LLVM's inliner, in which case \tool{}
makes all 12 of the benchmarks smaller with a total savings of about
15\,KB\@.

Performance of the SPEC benchmarks was also unevenly affected by
\tool{}: five become faster (perlbench, bzip2, hmmer, libquantum,
astar) and seven become slower (gcc, mcf, gobmk, sjeng, h264ref,
omnetpp, xalancbmk).
None of the differences was very large.
The overall SPECint\_base2006 rating was 36.3 without \tool{} and 36.0
with it (larger ratings are better).
Again, we would not expect introducing constants to reduce
performance; clearly, there are some interesting interactions between
\tool{} and the rest of LLVM's optimization pipeline.
We have not yet analyzed the situation further.
In general, the SPEC benchmarks have received a lot of attention from
compiler developers and they are not a place where we expect to find
much low-hanging fruit.

\section{Related Work}
\label{sec:related}


\paragraph{Superoptimization for LLVM and GCC}
A superoptimizer by Sands~\cite{Sands11} pointed out many opportunities
for optimizations that had not yet been implemented in LLVM around 2010
and 2011.
This tool had many similarities to \tool{}: it extracted directed
acyclic graphs of LLVM IR during compilation and then attempted to
optimize them, it focused on the integer subset of LLVM, and it
presented its results ordered by static profile count.
On the other hand, Sands' tool relied on testing to perform
equivalence checking and was, therefore, unsound; this did not lead to
miscompilations since it worked offline, and was only intended to
generate suggestions for developers to implement.
To synthesize right-hand sides, this superoptimizer applied various
heuristic simplifications of the left-hand side.
Analogously, the GNU superoptimizer~\cite{Granlund92} was an unsound,
enumeration-based tool that was used to improve GCC's code generation
by deriving ways to turn code with control flow into straight-line
code.

\paragraph{An online superoptimizer}
Most previous superoptimizers have been intended for offline use: it
has not been practical to run them during regular compiles.
Bansal and Aiken's tool~\cite{Bansal06} is one of the few exceptions:
it achieved good performance by caching its results in a database
(among other techniques).
Additionally, it was sound, using a SAT solver to verify equivalence.
This tool was a direct inspiration for \tool{}, though our work
improves upon it in important ways, for example by using synthesis
instead of enumeration to construct RHSs, by learning dataflow facts
from diverging and converging control flow, and by extracting
instructions that are related by dataflow rather than instructions
that happen to be close to each other in the instruction stream.
Bansal and Aiken's tool, on the other hand, could deal with memory
accesses and vector instructions, while \tool{} cannot.

\paragraph{The original superoptimizer}
Massalin~\cite{Massalin87} coined the term \emph{superoptimizer} to
emphasize the fact that regular optimizers produce code that is
far from optimal.
His tool was offline and unsound, using enumeration to discover
right-hand sides and testing to rule out inequivalent ones.

\paragraph{The earliest superoptimizers}
Fraser's 1979 paper~\cite{Fraser79} is the earliest work we know of
that captures all of the essential ideas of superoptimization:
extracting instruction sequences from real programs, searching for
cheaper ways to compute them, and checking each candidate using an
equivalence checker.
Moreover, unlike much subsequent work, Fraser's equivalence checker
was sound.
Subsequently, in 1984, Kessler~\cite{Kessler84} developed a sound
superoptimizer for a Lisp compiler.

\paragraph{Recent superoptimizers}
There has been significant progress in superoptimizers in recent
years.
Optgen~\cite{Buchwald15}, like \tool{} and unlike almost all of the
other work described in this section, operates on IR rather than
assembly language.
It generates all optimizations up to a cost limit, rather than
extracting code sequences from programs.
Unlike \tool{}, it can synthesize optimizations containing symbolic
constants.
STOKE~\cite{Schkufza13,Schkufza14,Sharma15} is a sound superoptimizer
that uses randomized search, rather than solver-based synthesis, to
find cheap right-hand sides.
It can handle many features that \tool{} cannot, including memory,
floating point instructions, approximations, and even, in conjunction
with the DDEC tool~\cite{Sharma13}, loops.
Chlorophyll~\cite{Phothilimthana14} is a synthesis-aided compiler for
a 144-core chip.
Finally, GreenThumb~\cite{Phothilimthana16} is a generic
superoptimizer that can be easily retargeted; it has been used to
optimize LLVM IR\@.

\paragraph{Synthesis}
There has been enormous progress in program sketching and synthesis in
the last few years.
Our work mainly builds upon a single paper, Gulwani et
al.~\cite{Gulwani11}.

\section{Future Work}

Every optimization discovered by \tool{} is specific: it is not
parameterized by bitwidths, values of constants, or choice or ordering
of instructions.
On the other hand, optimizations implemented by compiler developers are
almost always generic along one or more of these dimensions.
It would be useful for \tool{} to automate some of this
generalization, for example emitting an Alive~\cite{Lopes15} pattern
for each optimization it discovers.
Generalization would improve our rankings by preventing important
optimizations that happen to have many different forms from hiding low
in the rankings where they are unlikely to be looked at.
Generalization would also present compiler developers with
optimizations that are more like the ones that they will presumably
implement, saving them the effort of generalization by hand.
In particular, it can be very difficult to write a sound and precise
precondition for an optimization, especially in the presence
of undefined behavior.

A middle-end superoptimizer cannot exploit
target-specific code sequences.
Our hypothesis is that future compilers should contain two
superoptimizers.
First, one in the middle-end that, like \tool{}, tries to generate
constants and perform other obviously-profitable transformations that
can be exploited by other middle-end passes.
Second, a target-aware backend superoptimizer that interacts with
instruction selection, scheduling, and register allocation.

Finally, we believe that \tool{} IR can be extracted from CompCert RTL
in the same way that we extract it from LLVM IR and Microsoft Visual
C++ IR\@.
Synthesizing optimizations is then trivial.
The research problem is to take the resulting equivalence proofs
emitted by a proof-producing solver and integrate them into CompCert's
overall proof.

\section{Conclusion}

We created a synthesizing superoptimizer that is integrated with LLVM,
and we showed that it can derive many optimizations that LLVM does not
yet have.
As a result of our work, some of these optimizations have been
implemented by hand in LLVM, where they benefit all users of that
toolchain.
We also showed that when \tool{} is used to synthesize integer
constants, it can be used as an online compiler that results in a
significant code size reduction in some cases.
Finally, we showed that \tool's IR is generic enough that Microsoft
Visual C++'s IR can be translated into it, and that \tool's
suggestions about missing optimizations were actionable by the MSVC
compiler developers.

\bibliography{bib}


\bibliographystyle{abbrvnat}

\end{document}